# Wafer-scale Synthesis of Mithrene and its Application in 2D Heterostructure UV Photodetectors

**Maryam Mohammadi[1,]\*, Stefanie L. Stoll[1], Analía F. Herrero[2], Sana Khan[1,3], Federico Fabrizi[1,3], Christian Gollwitzer[2], Zhenxing Wang[1], Surendra B. Anantharaman [1, 4,]\*, Max C. Lemme[1,2,]\***

[1]AMO GmbH, Otto-Blumenthal-Straße 25, Aachen, 52074, Germany

[2]Physikalisch-Technische Bundesanstalt, Abbestraße 2–12, Berlin, 10587, Germany

[3]Chair of Electronic Devices, RWTH Aachen University, Otto-Blumenthal-Straße 25, Aachen, 52074, Germany

[4]Low-dimensional Semiconductors Lab, Department of Metallurgical and Materials Engineering, Indian Institute of Technology Madras, Chennai, 600036, India

\*Corresponding Authors: mohammadi@amo.de, sba@smail.iitm.ac.in, max.lemme@eld.rwth-aachen.de

**Abstract**

Silver phenylselenide (AgSePh), known as mithrene, is a two-dimensional (2D) organic-inorganic chalcogenide (MOC) semiconductor with a wide direct band gap, narrow blue emission and in-plane anisotropy. However, its application in next-generation optoelectronics is limited by crystal size and orientation, as well as challenges in large-area growth. Here, we introduce a controlled tarnishing step on the silver surface prior to the solid-vapor-phase chemical transformation into AgSePh thin films. Mithrene thin films were prepared through thermally assisted conversion (TAC) at 100°C, incorporating a pre-tarnishing water ($H_2O$) vapor pulse and propylamine ($PrNH_2$) as a coordinating ligand to modulate $Ag^+$ ion reactivity and facilitate the conversion of $Ph_2Se_2$ into an active intermediate. The AgSePh thin films were characterized by X-ray diffraction (XRD), scanning electron microscopy (SEM), and grazing incidence wide-angle X-ray scattering (GIWAXS). The pre-tarnishing process, combined with organic ligands, resulted in large crystals exceeding 1 µm and improved homogeneous in-plane orientation, while also enabling the selective, wafer-scale synthesis of mithrene on 100 mm wafers. Furthermore, the films were integrated on planar graphene field-effect phototransistors (GFETs) and demonstrated photoresponsivity beyond 100 A/W at 450 nm, highlighting mithrene's potential for blue light-detection applications.

**Introduction**

Silicon photodetectors (PDs) provide high sensitivity, fast response, low power consumption, and low fabrication costs due to the established silicon semiconductor mass-production technology [1]. However, silicon's weak response to ultraviolet (UV) and blue light, along with the large dark current, are critical drawbacks for practical applications in the UV-blue region [2–5]. Most commercially available UV-blue PDs are based on wide-bandgap inorganic semiconductors, such as silicon carbide, gallium nitride (GaN), and gallium phosphide (GaP). Fabricating inorganic semiconductor PDs involves complex processes, such as molecular beam epitaxy (MBE) and metal-organic vapor phase epitaxy (MOVPE), which require high-temperature fabrication techniques [6–8]. Therefore, developing semiconductor materials with low-temperature processes for UV-blue light detection is of great interest, because they can be easily integrated in the back-end-of-line (BEOL) of silicon complementary metal oxide semiconductor (CMOS) technology.

2D metal-organic chalcogenides (MOCs) are an emerging class of hybrid organic-inorganic 2D semiconductors with a chemical formula of $[M(ER)]_n$, where M = copper (Cu(I)), silver (Ag(I)), and gold (Au(I)); E = sulfur (S), selenium (Se), and tellurium (Te); and R is an organic hydrocarbon [9]. 2D MOCs crystallize as three-dimensional solids of 2D layers bound together by interlayer van der Waals forces, similar to transition metal dichalcogenides (TMDs) and 2D perovskites (Figure 1a). Each 2D layer consists of an inorganic sheet sandwiched between organic ligands, which are covalently bonded to the inorganic sheet [10–13]. The presence of organic ligands prevents electronic interactions between inorganic layers, distinguishing 2D MOCs from TMDs, which exhibit layer-dependent electronic properties [14–17]. Furthermore, 2D MOCs differ from 2D perovskites in the covalent nature of bonding between organic and inorganic components, which provides chemical stability in air and solvents [18,19]. The 2D MOC silver phenylselenolate (AgSePh), also known as mithrene, has attracted considerable attention owing to its wide and direct band gap, low-temperature growth (≤100°C), ultrastrong light–matter coupling [20], capability for UV-blue light absorption [21], narrow blue (~467 nm) emission [22] which makes them potentially compatible with BEOL fabrication processes [23,24]. Additionally, the bandgap and, hence, the light absorption of mithrene can be tuned by modifying its organic ligands, further enhancing its potential for optoelectronic applications [10,22]. Therefore, mithrene is a promising lead-free alternative for light detection applications in the UV–blue range [13].

Despite mithrene's many advantages, there remains a need for readily controllable synthesis method to achieve high-quality, preferentially crystal-oriented mithrene thin films. Standard solid-vapor-phase methods for MOCs involve oxygen ($O_2$) plasma and UV-assisted tarnishing techniques [6,25] to form unstable silver complex components, which react with diphenyl diselenide ($Ph_2Se_2$) gas to produce mithrene. These methods lack precise control over reaction parameters, which has hindered the control of the lateral size of mithrene crystals and the crystal orientations in thin films.

Here, we introduce a thermally assisted conversion (TAC) technique for the scalable growth of mithrene at 100°C. The technique induces the formation of silver complexes through gentle water ($H_2O$) vapor pulses and the addition of propylamine ($PrNH_2$) organic ligands during the growth process (Figure 1b). This approach improves the lateral crystal size and orientation while simultaneously increasing photoluminescence intensity. The organic ligands play two roles: promoting the formation of a key reactive intermediate and slowing the conversion reaction to AgSePh, which leads to higher-quality crystals[10] than current state-of-the-art methods. We scale the method to 100 mm wafers and demonstrate 2D mithrene / graphene heterostructure PDs operating in the UV wavelength range.

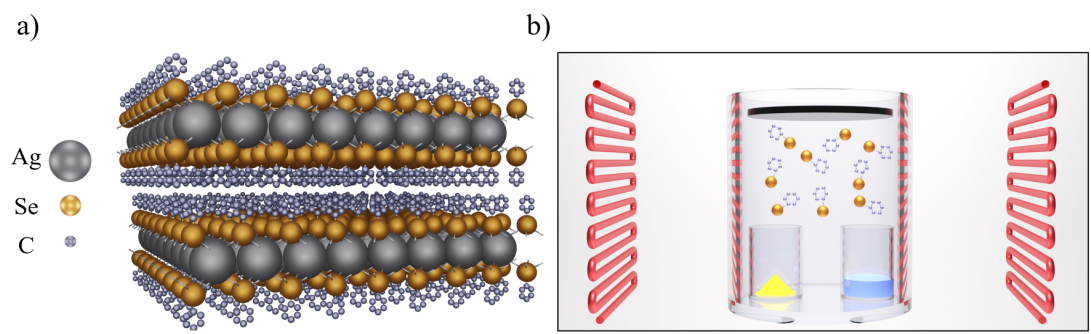

**Figure 1.** (a) Schematic of an AgSePh layered nanostructure represented at the molecular level. (b) Schematic of the TAC process. The entire process happens in a reaction vial in an oven.

**Results and Discussion**
**Synthesis of AgSePh thin films by amine addition**

A solid–vapor-phase chemical transformation method was employed to convert Ag films sputtered on $Si/SiO_2$ or bare glass substrates into AgSePh (details of the materials and processes can be found in the Methods section). AgSePh thin films were synthesized using two different pre-tarnish methods. Tarnishing reactions were performed by exposing a 10 nm sputtered silver thin film to mild $O_2$ plasma treatment (100 W) [25] or to $H_2O$ vapor pulse treatment. DMSO vapor was also selected as an assisting agent to control the size and orientation of AgSePh crystals because of its ability to form complexes with silver ions [13]. We propose the following conversion processes:

- $O_2$ plasma treatment

After exposure to mild $O_2$ plasma (100 W), silver oxidizes to $Ag_2O_2$, which is a 1:1 molar mixture of silver(I) oxide ($Ag_2O$) and silver(III) oxide ($Ag_2O_3$)[25,26]. The stability of silver oxides decreases as the oxidation state increases, leading to thermal decomposition at high temperatures into thermodynamically stable $Ag_2O$ [25]. $Ag_2O$ can react with gaseous $Ph_2Se_2$ at 100°C, resulting in the formation of yellowish mithrene crystals. SEM images (Figure 2a-e) and XRD patterns (Figure 2f) confirm the complete conversion of silver to mithrene in all $O_2$ plasma-treated samples, while residual Ag peaks are observed in the control sample without pre-treatment (Figure 2g). The characteristic crystal structure of mithrene is evident from the diffraction peaks associated with the (002), (004), and (006) crystallographic planes, observed at 2θ angles of approximately 6°, 12°, and 18°, respectively. This series of first- and higher-order reflections corresponds to an interlayer spacing (d) of 1.4 ± 0.03 nm, consistent with those noted in mithrene synthesized via both biphasic synthesis and vapor-phase methods[10,13]. With increasing $O_2$ plasma exposure time, the X-ray diffraction peak intensity decreased (Figure 2f). XRD, photoluminescence spectroscopy, and AFM results indicated that the film made with a 30-second $O_2$ plasma treatment has more pronounced crystallinity (Figure 2f), stronger luminescence at 467 nm (Figure S1), and the lowest RMS roughness of 31.5 nm (Figure S2). SEM (Figure 2b-e). and AFM (Figure S2) results show irregular mithrene crystals formed through the $O_2$ plasma treatment process.

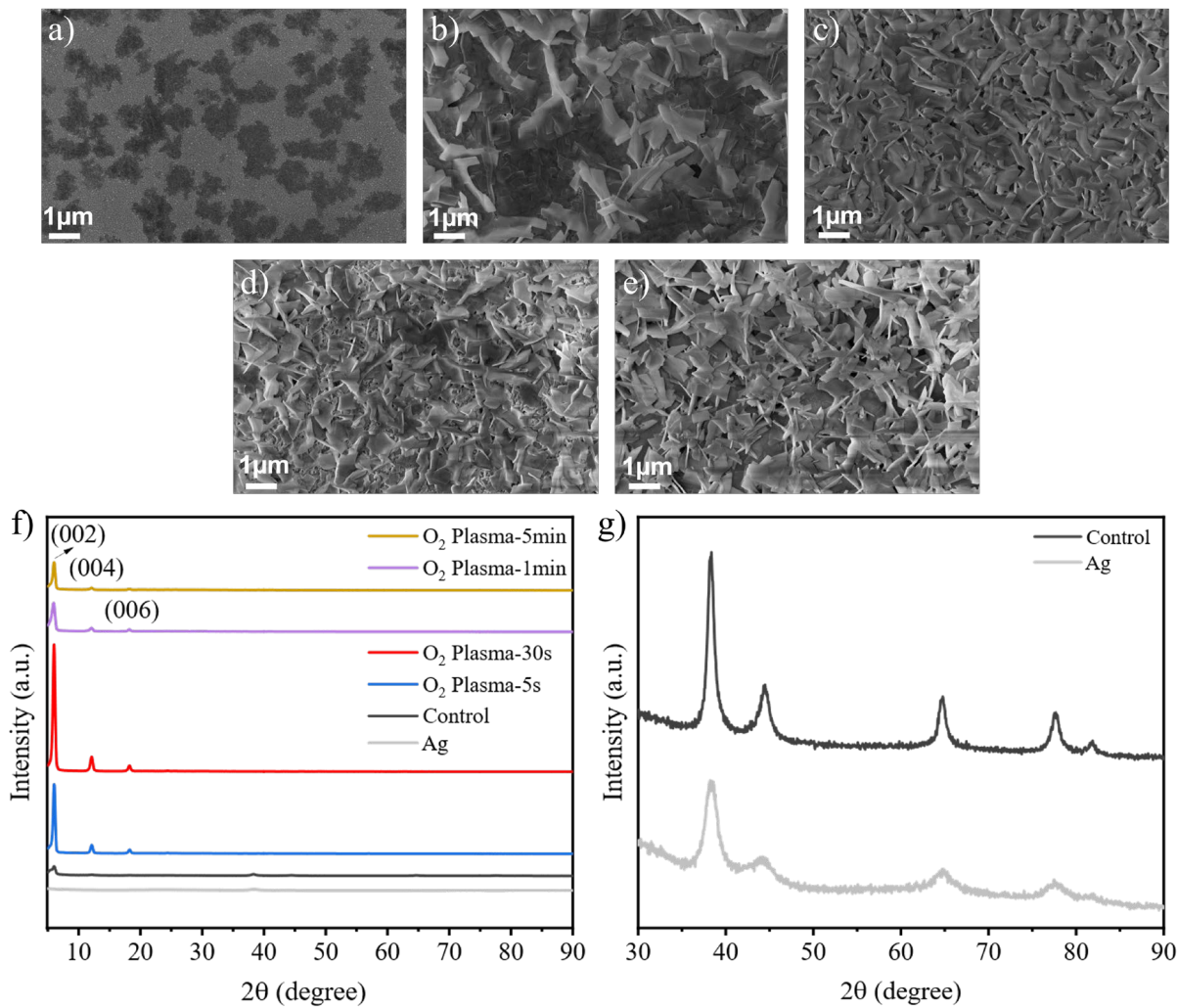

**Figure 2.** SEM images of $O_2$ plasma-treated samples at different treatment durations: (a) 0 seconds, (b) 5 seconds, (c) 30 seconds, (d) 1 minute, (e) 5 minutes. f) XRD patterns of $O_2$ plasma-treated samples. (g) Zoomed-in XRD patterns for Ag and control samples.

- $H_2O$ vapor pulse treatment

Silver, like other metals, attracts a thin layer of water during the water vapor pulse treatment. This water layer acts as an electrochemical electrolyte, enabling corrosion in a sulfur-containing atmosphere, which can form from the oxidation of DMSO in the reaction vial or from atmospheric carbonyl sulfide (0.5 ppb) [11]. This process facilitates the subsequent dissolution of solid silver and the formation of mithrene[27]. The XRD pattern of AgSePh prepared via $H_2O$ vapor pulse treatment shows a significant enhancement in the intensity of the (002) peak, which is related to the increase in crystal size and improved crystallinity (Figure 3a). The SEM image of the AgSePh thin film (Figure 3b) confirms the in-plane crystal growth of mithrene, with the extension in the lateral direction significantly impacted. During $O_2$ plasma treatment, the conversion of the silver layer to silver oxides, which have a lower density, results in a porous surface with an enhanced number of reaction sites for $Ph_2Se_2$ gases (Figure

S3a). In contrast, water vapor exposure maintains the silver surface topography as nearly dense (Figure S3b) and slows down crystal growth, allowing for the formation of higher-quality crystals. This is consistent with the SEM images in Figure S4, which illustrate that the reaction rate of $H_2O$ vapor pulse-treated silver with $Ph_2Se_2$ gas is much slower than that of the 30-second $O_2$ plasma-treated silver under the same reaction time.

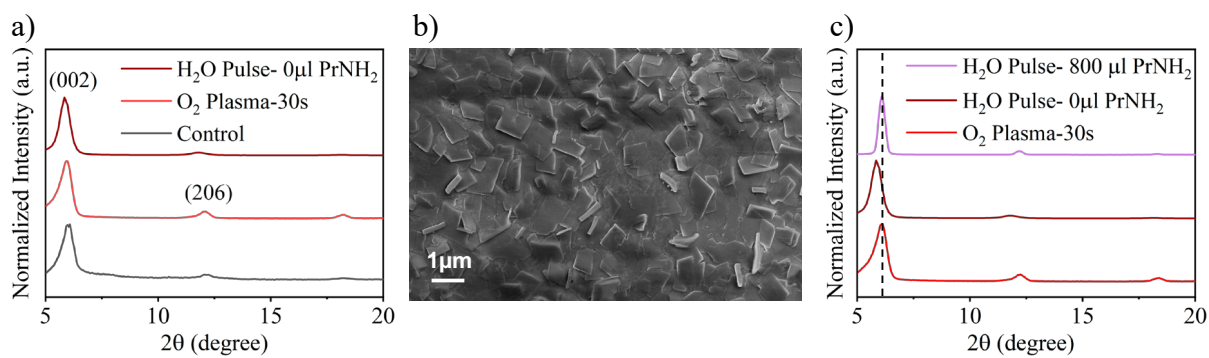

**Figure 3.** (a) XRD patterns of the control and treated samples. (b) SEM image of the $H_2O$ vapor pulse-treated sample. (c) XRD patterns of control, $O_2$ pulse-treated, $H_2O$ vapor pulse-treated and $H_2O$ vapor pulse-treated samples modified with 800 μl of organic $PrNH_2$ ligand.

A comparison of the XRD results of the mithrene layers obtained from the $O_2$ plasma and $H_2O$ vapor pulse treatment methods revealed a slight shift to lower angles in the XRD peaks for the $H_2O$ vapor pulse treatment method, which could be related to strain in the crystal lattice (Figure 3c). Using additives to control the crystallization dynamics is a strategy for regulating lattice strain in the mithrene crystals and passivating surface defects[28]. Additionally, it has been reported that the formation of silver-amine complexes reduces the mithrene growth rate. [29–32]. To leverage the benefits of the amine functional group in the synthesis of 2D mithrene crystals, we added $PrNH_2$ during the solid-vapor-phase growth process, varying the amount to 200 μl, 400 μl, and 800 μl. The XRD results in Figure S5 indicate that the amines did not alter the product of the conversion reaction. Figure 3c show that the addition of $PrNH_2$ during the growth reaction can regulate the strain in AgSePh crystals. However, increasing the amine concentration resulted in the formation of more laterally oriented crystals and a smooth film surface. SEM (Figure 4a-c) and AFM (Figure S6) images revealed a smooth, featureless basal surface attributed to the (002) crystallographic plane. Figure S7 compares the photoluminescence of AgSePh thin films prepared by $O_2$ plasma and $H_2O$ vapor pulse treatments, with and without 800 μl $PrNH_2$. With $H_2O$ vapor pulse treatment with and without organic ligand, the PL intensity increased, while the emission peak position remained almost unchanged, with only a 1 nm variation. This is consistent with the findings of Schriber et al., which indicate that the luminescence maximum wavelength ($\lambda_{max}$) has almost no correlation with mithrene's crystal size or morphology [33]. The further enhancement in PL intensity for the water vapor pulse treatment with 800 µl PrNH$_2$ may be associated with effective surface passivation. All resulting microcrystals exhibited uniform blue luminescence, with an emission wavelength at 466 nm.

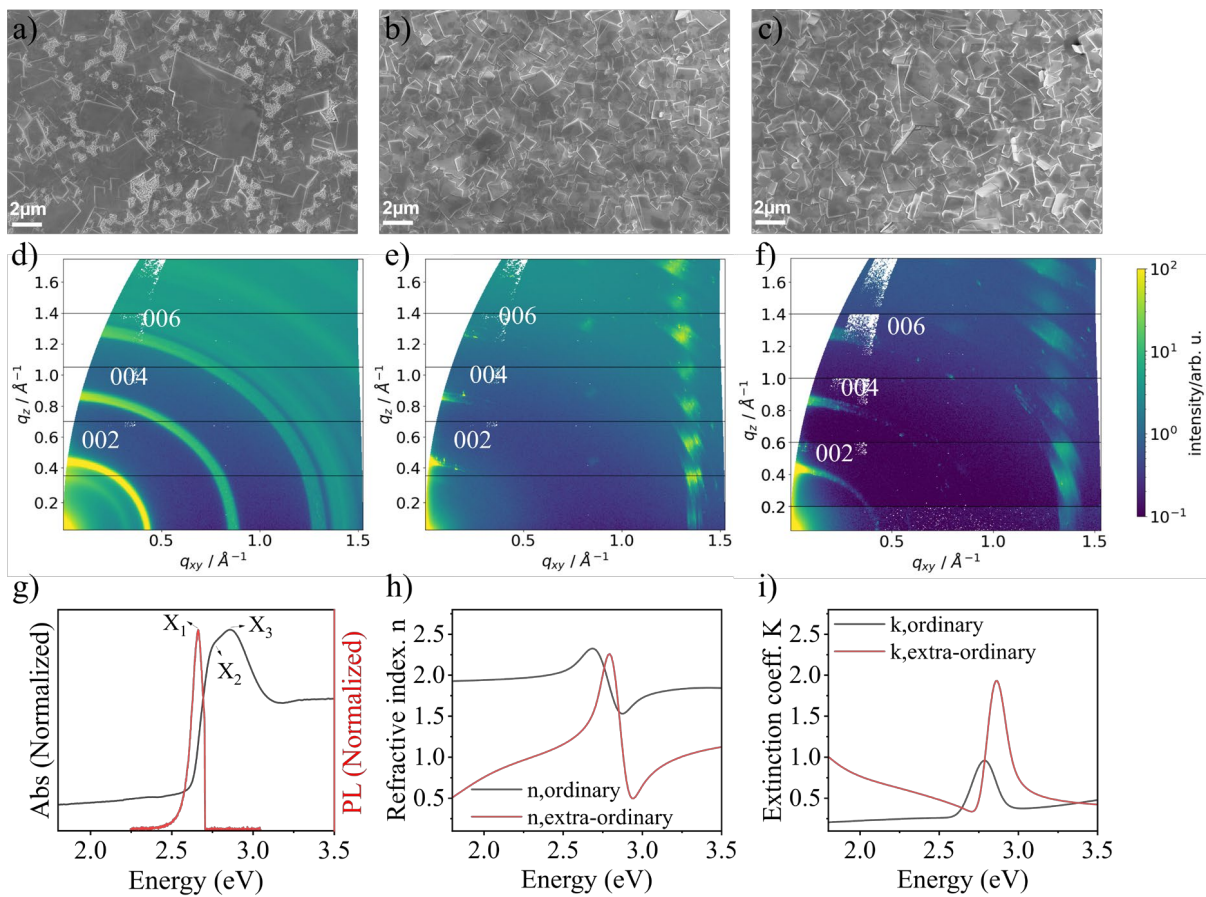

**Figure 4.** SEM images of H$_2$O vapor pulse-treated samples with (a) 200 µL PrNH$_2$, (b) 400 µL PrNH$_2$, (c) 800 µL PrNH$_2$. 2D GIWAXS images of the (d) O$_2$ plasma-treated sample, (e) H$_2$O vapor pulse-treated sample, and (f) H$_2$O vapor pulse-treated sample with 800 µl of PrNH$_2$. The intensity scale is the same for all the images. (g) Absorbance (black line) and photoluminescence (PL) spectra (red line), (h) Optical extinction coefficient k for the in-plane and out-of-plane components, (i) Optical refractive index n for the in-plane and out-of-plane components.

The orientation distribution in the O$_2$ plasma-treated and H$_2$O vapor pulse-treated films, both without and with 800 µl of PrNH$_2$, was evaluated via GIWAXS, showing different features in the scattering patterns. Debye-Scherrer rings, that indicate a random orientation on powder-like films are observed for the O$_2$ plasma-treated film (Figure 4d). The Bragg scattering peaks of crystals grown with H$_2$O vapor pulse treatment, both without and with 800 µl of PrNH$_2$, demonstrate a strongly preferential orientation (Figure 4e and 4f). The high-intensity peaks in the q$_z$ coordinate plane, corresponding to the (002), (004), and (006) crystallographic planes of the mithrene unit cell, are identified in Figures 4e and 4f. This shows a preferred face-on orientation of the mithrene crystallites in the films, where the (001) face is parallel to the substrate. Low intense out-of-plane scattering peaks (at higher q$_{xy}$) are the result of non-uniformity in the orientation. However, the results show that the addition of PrNH$_2$ decreases the intensity of those contributions (Figure 4f).

The spectra in Figure 4g show the absorption and photoluminescence of the mithrene treated with H$_2$O vapor pulses and synthesized with 800 µl of PrNH$_2$. The absorbance spectrum (black line) shows three excitonic resonances at $X_1$ = 2.66 eV, $X_2$ = 2.76 eV, and $X_3$ = 2.87 eV. Photoluminescence emission (red line) occurs as a single peak at 2.66 eV (~ 466 nm), which corresponds to the lowest excitonic resonance $X_1$. Considering that the layered structure leads to high in-plane charge carrier confinement, significant polarization anisotropy is expected for excitonic optical transitions. To test this hypothesis, we performed spectroscopic ellipsometry using a J.A. Woollam M-2000 spectroscopic ellipsometer to evaluate the degree of anisotropy in the mithrene films. The data obtained were analyzed and modeled via the J.A. Woollam CompleteEASE software package. The best fit to the measured ellipsometry data was achieved with a film thickness of 100 nm, assuming uniaxial anisotropy. The model included two B-spline layers to represent the optical constants in out-of-plane (ordinary) and in-plane (extra-ordinary) directions. These B-spline layers were further parameterized via Gaussian oscillators to capture the observed excitonic and band edge-like features accurately. The extracted optical refractive indices and extinction coefficients from these fits are plotted in Figures 4h and i. The observed anisotropy in the optical refractive index and extinction coefficients, despite the polycrystalline nature of the measured area, is attributed to a strong (001) preferential orientation of the crystalline domains in the film, as seen in GIWAXS (Figure 4f). The magnitude of the extinction coefficient in the in-plane direction is noticeably greater.

Mithrene thin films were also grown on a 100 mm wafer (Figure 5a) and selectively on a chip with predefined silver structures (Figures 5b and c). The refractive index map for mithrene growth on 100 mm wafer is shown in Figure S8. The thickness across the wafer ranged from ~100-140 nm (Figure S8a). The average ordinary and extraordinary refractive indices at 1.96 eV (632 nm) closely matched those obtained from a single-spot measurement, 1.90 and 0.71, respectively (Figures S8b and c), demonstrating a uniform in-plane refractive index distribution.

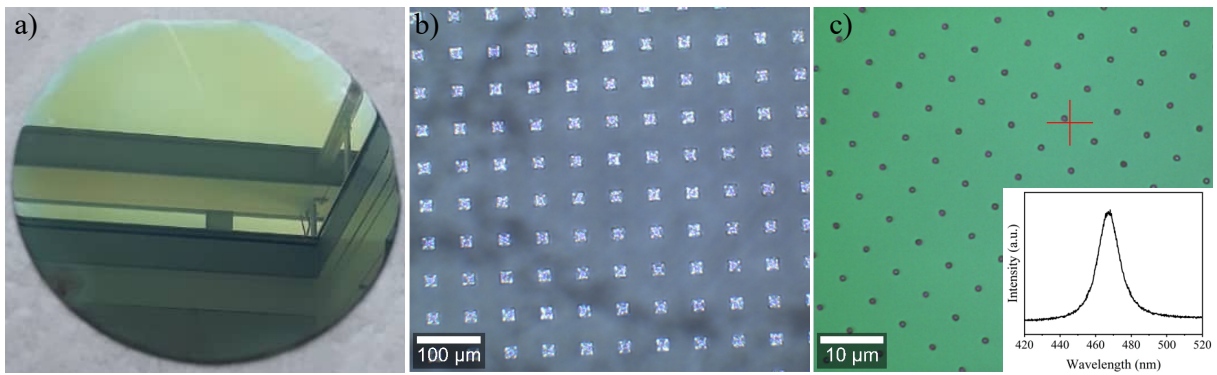

**Figure 5.** (a) Wafer-scale growth of mithrene on a 100 mm wafer. (b, c) Selective mithrene growth; the inset shows the PL at 467 nm from the point marked with a red cross.

**Photodetectors**

We fabricated PDs in a semiconductor process flow (see Methods section). A schematic of the GFET/mithrene PD and a top-view photograph are shown in Figures 6a and b. We performed Raman spectroscopy to compare the graphene quality before and after silver deposition (Figures S9). The morphology of the mithrene film grown on graphene in the active device area is shown in Figure 6c. The SEM confirms the in-plane crystal growth of mithrene, with crystals sizes exceeding 1 μm.

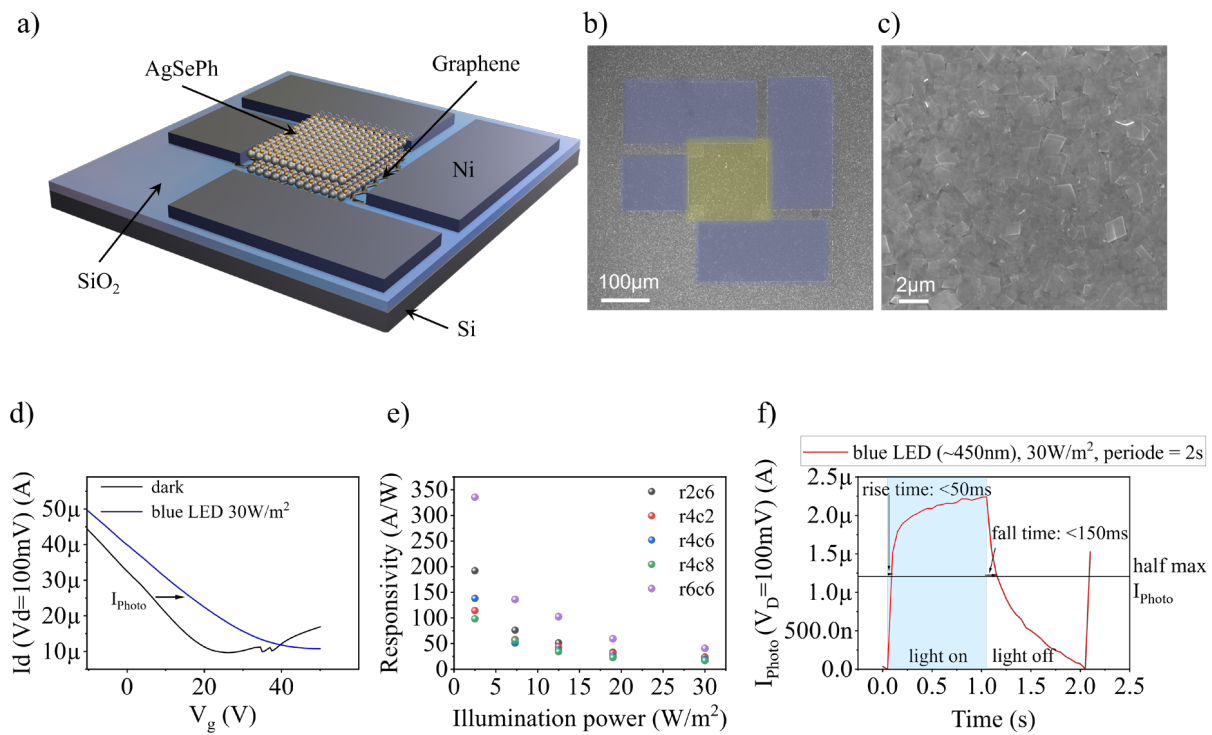

**Figure 6**. (a) Schematic of the GFET/mithrene device. (b) Top-view photograph of the device. (c) Morphology of the mithrene film on the device's active area. (d) Transfer characteristics of a Mithrene/Graphene field effect transistor under dark conditions and with blue light illumination (at ~450 nm with an illumination power of 30 W/m$^2$). (e) Responsivity of five devices at different illumination powers (The legend shows the names of the devices). (f) Time-resolved drain current of an example device reacting to one switching cycle (period length of 2 s) of a blue (~450 nm) LED with an illumination power of 30 W/m$^2$.

We measured the transfer characteristics of the devices by applying a constant source-drain voltage of 100 mV, while varying the gate voltage to the sample substrate via the chuck (Figure 6d). The minimum point of the transfer characteristic (charge neutrality point) [34] shifted to higher voltages under blue illumination, suggesting the injection of holes into the graphene during illumination [35]. The responsivities of various devices are shown in Figure 6e as a function of different illumination powers, exceeding 1 A/W due to the photogating effect. We speculate that under illumination, holes are introduced into the graphene, while electrons remain in the mithrene layer. Due to the high mobility of charge carriers in the graphene layer, these carriers can circulate multiple times during the lifetime of the trapped charge carriers in the mithrene layer. Consequently, the remaining charge carriers in the mithrene layer function as a secondary gate (photogating), further enhancing the photocurrent [36]. However, at lower illumination powers, the extended average lifetime of the trapped charges [37–39] results in an increase in the responsivity of the devices. The trend of increasing responsivity with decreasing illumination power shows no signs of saturation, indicating that the maximum responsivity has not yet been reached [35].

The time-resolved response to changes in illumination was characterized. Figure 6f shows the current through the device in response to a single switching cycle of the blue LED with an illumination power of 30 W/m$^2$ and a period length of 2 s for a representative device. The horizontal line indicates half of the maximum photocurrent after the light is switched on, with a rise time of less than 50 ms and a fall time of less than 150 ms. The measurement setup has a minimal temporal resolution of 50 ms, which means that both the rise and fall times are overestimated in this measurement. The rise time of heterostructure PDs is dominated by the time it takes for the charge carriers to reach the graphene surface, whereas the fall time is influenced by the charges remaining trapped in the functionalization layer. These response times are comparable to those observed in other graphene-hybrid phototransistors[35].

**Conclusions**

In summary, we demonstrate wafer-scale (100 mm) mithrene growth for the first time using a solid-vapor-phase chemical transformation process of metallic silver to mithrene (AgSePh) with controlled pre-tarnishing by $H_2O$ vapor pulse treatment. The $H_2O$ vapor pulses, combined with $PrNH_2$ as an organic coordinating ligand, reduced the reactivity of $Ag^+$ ions and slowed the conversion process, yielding flatter, larger crystals exceeding 1 μm with (001) preferred orientation and a high in-plane extinction coefficient. The mithrene films were integrated into GFETs without significantly damaging the graphene, as evidenced by Raman spectroscopy. Furthermore, we demonstrated Mithrene/GFET PDs based on charge carrier transfer into the graphene. The PDs showed responsivities beyond 100 A/W at 450 nm through a photogating mechanism, validating mithrene's reliability as an air-stable photoactive layer.

**Materials and methods**

**Materials.** Silver pellets (Ag, 99.99% pure) were obtained from EVOCHEM. Diphenyl diselenide ($Ph_2Se_2$, 98%), dimethyl sulfoxide (DMSO, anhydrous, ≥ 99.9%), and propylamine ($PrNH_2$, ≥99%) were received from Sigma-Aldrich and used without further purification.

**Mithrene thin films.** We prepared two sets of samples for our experiments: $Si/SiO_2$ substrates and bare glass substrates with dimensions of 20 mm × 20 mm were first sonicated for 10 minutes in acetone and isopropyl alcohol. Then, they were blow-dried with nitrogen ($N_2$) and exposed to an oxygen ($O_2$) plasma in a Tepla Semi 300 tool for 5 minutes. Ag films with thicknesses of approximately 10 nm were sputtered onto substrates in a Creavac Creamet 500 S2 at $5.3 \times 10^{-7}$ mbar with a deposition rate of ~1.8 Å/s. The Ag layers were pre-tarnished with mild $O_2$ plasma at a nominal power of 100 W [25] with exposure times of 1 second, 30 seconds, 1 minute, and 5 minutes. A second set of samples for pre-tarnished silver films were prepared by exposing the samples to 15 seconds of $H_2O$ vapor pulses at 50°C in an atomic layer deposition tool (Oxford Instruments, FlexALRPT). The tarnished silver films, 60 mg of $Ph_2Se_2$ powder, and 200 μL of DMSO were sealed inside a reaction vial and heated to 100°C in an oven to produce AgSePh films. The amount of the organic ligand was varied by adding 200 μL, 400 μL, or 800 μL of $PrNH_2$ to the sealed reaction vial to study its effect on the mithrene crystal size and orientation.

The mithrene thin films were characterized by X-ray diffraction (XRD) measurements, scanning electron microscopy (SEM), atomic force microscopy (AFM), UV-visible (UV-vis) and photoluminescence (PL) spectroscopy, grazing-incidence wide-angle X-ray scattering (GIWAXS), and spectroscopic ellipsometry. XRD with filtered Cu-Kα radiation with a wavelength of 1.5405 Å was conducted using a PANalytical instrument at a current of 40 mA and a voltage of 40 kV. SEM measurements were performed with a Zeiss SUPRA 60 at 4 kV with a working distance of 4 mm. AFM images were captured from each sample using an atomic force microscope (Bruker Dimension Icon) in tapping mode. UV-vis spectroscopy of the mithrene thin films on glass substrates was performed with a PerkinElmer Lambda 1050 spectrophotometer. Photoluminescence images were taken with a WiTec confocal Raman microscope by scanning the sample with a 405 nm continuous-wave laser at 2 μW power. Spectroscopic ellipsometry was performed using a J. A. Woollam M-2000 spectroscopic ellipsometer, with data analyzed and modeled via the J. A. Woollam CompleteEASE software package.

Automated refractive index mapping on 100 mm wafer was conducted using a Philips PQ Ruby Ellipsometer at 632 nm. GIWAXS measurements were performed at the FCM beamline of the Physikalisch-Technische Bundesanstalt laboratory at the BESSY II storage ring. The FCM beamline can deliver monochromatic radiation between 1.7 keV and 10 keV with moderate flux[40], and features slits and pinholes for beam shaping. For this experiment, a beam spot of less than 0.5 mm × 0.5 mm with minimal halo was achieved at the sample position using a pinhole pair. The sample and PILATUS detector were located within a vacuum sample chamber allowing for three-dimensional positioning and rotation of the sample. A single-photon counting PILATUS 100k detector[41] with negligible dark signal was positioned on a concentric arm approximately 200 mm from the chamber's center of rotation. The incoming photon energy was set to 8 keV, and the samples were illuminated under a grazing-incidence angle of 0.2°. For each sample, 5 overlapping scattering images were acquired covering an angular range of 20°. The acquisition time of each image varied depending on the scattered angles and then normalized.

**PD fabrication.** We fabricated mithrene/graphene PDs on Si/SiO$_2$ substrates. The PDs were based on graphene field effect transistors (GFETs), where thermally oxidized p-doped silicon substrates with 90 nm SiO$_2$ served as a global back gate. The graphene was transferred onto the substrates using a polymer-assisted wet transfer method, and patterned with AZ5214-E photoresist, followed by reactive ion etching in an O$_2$ plasma. Source and drain contacts were fabricated by thermal evaporation of 50 nm of nickel (Ni) onto a photoresist mask, followed by a lift-off process. Next, microcrystalline mithrene thin films were prepared by sputtering 10 nm thick silver films onto the samples. We performed Raman spectroscopy on test samples to verify that the sputtering process did not significantly damage the graphene. The samples were then exposed to H$_2$O vapor pulses to induce pre-tarnishing. According to the analytical experiments, this specific process was the most promising route. The chips were then placed next to 60 mg of solid Ph$_2$Se$_2$, 200 μL of DMSO, and 800 μL of PrNH$_2$ for conversion to mithrene.

The PDs were characterized in a Lakeshore Cascade Summit 12000 Probe Station with a Hewlett Packard Precision semiconductor parameter analyzer A4156A under dark conditions and illumination from a blue light emitting diode (LED) with a wavelength of 450 nm at different power levels. The responsivity (R (A/W)) was calculated using the formula R=I$_{Photo}$/P$_{Illumination}$×A, where I$_{photo}$ (A), P$_{Illumination}$ (W/m$^2$), and A (1.6×10$^{-8}$m$^2$) represent photocurrent, illumination power density, and illuminated photoactive area, respectively [42]. The photocurrent is defined as the difference between the current under illumination and in dark conditions.

**Acknowledgements**

This project was funded by the Deutsche Forschungsgemeinschaft (DFG) within TRR 404 Active-3D (project number: 528378584) and Hiper-Lase (project numbers: GI 1145/4-1 and LE 2440/12-1); the European Union's Horizon 2020 research and innovation programme under the projects FOXES (951774), PERSEPHONe (956270), and MISEL (101016734); and the German Ministry of Education and Research (BMBF) through the project NEPOMUQ (13N17112). The authors thank P. Grewe and Dr. U. Böttger (Electronic Material Research Lab, RWTH Aachen University) for their support in the XRD measurements.

# Wafer-scale Synthesis of Mithrene and its Application in 2D Heterostructure UV Photodetectors

**Maryam Mohammadi[1],\*, Stefanie L. Stoll[1], Analía F. Herrero[2], Sana Khan[1,3], Federico Fabrizi[1,3], Christian Gollwitzer[2], Zhenxing Wang[1], Surendra B. Anantharaman [1, 4],\*, Max C. Lemme[1,2],\***

[1]AMO GmbH, Otto-Blumenthal-Straße 25, Aachen, 52074, Germany

[2]Physikalisch-Technische Bundesanstalt, Abbestraße 2–12 , Berlin, 10587, Germany

[3]Chair of Electronic Devices, RWTH Aachen University, Otto-Blumenthal-Straße 25, Aachen, 52074, Germany

[4]Low-dimensional Semiconductors Lab, Department of Metallurgical and Materials Engineering, Indian Institute of Technology Madras, Chennai, 600036, India

\*Corresponding Authors: mohammadi@amo.de, sba@smail.iitm.ac.in, max.lemme@eld.rwth-aachen.de

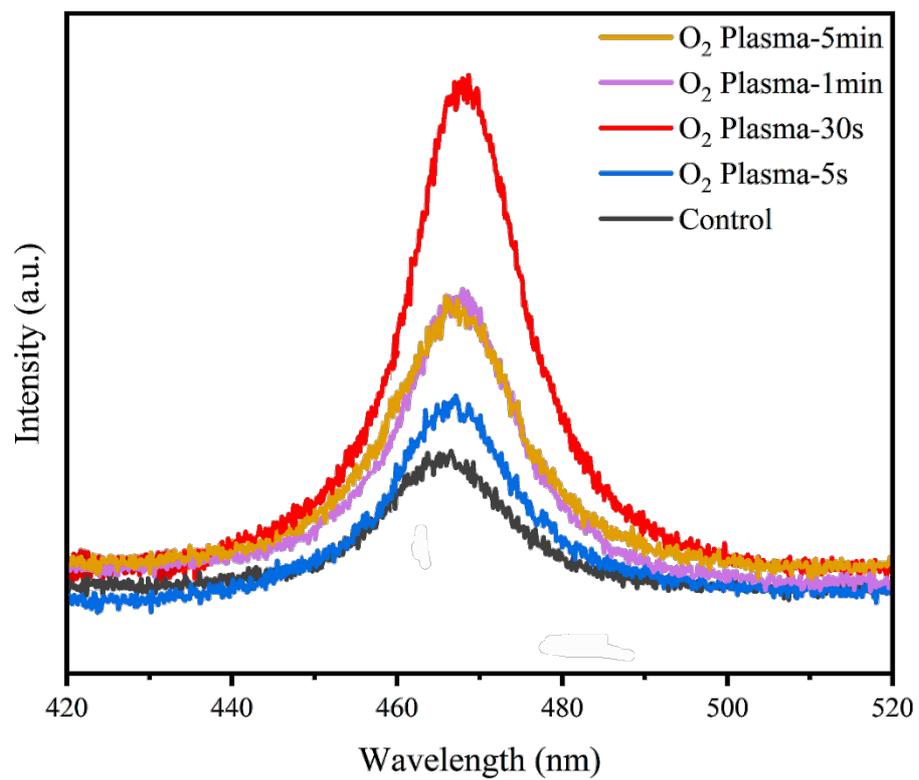

Figure S1. Photoluminescence spectra of O₂ plasma-treated samples.

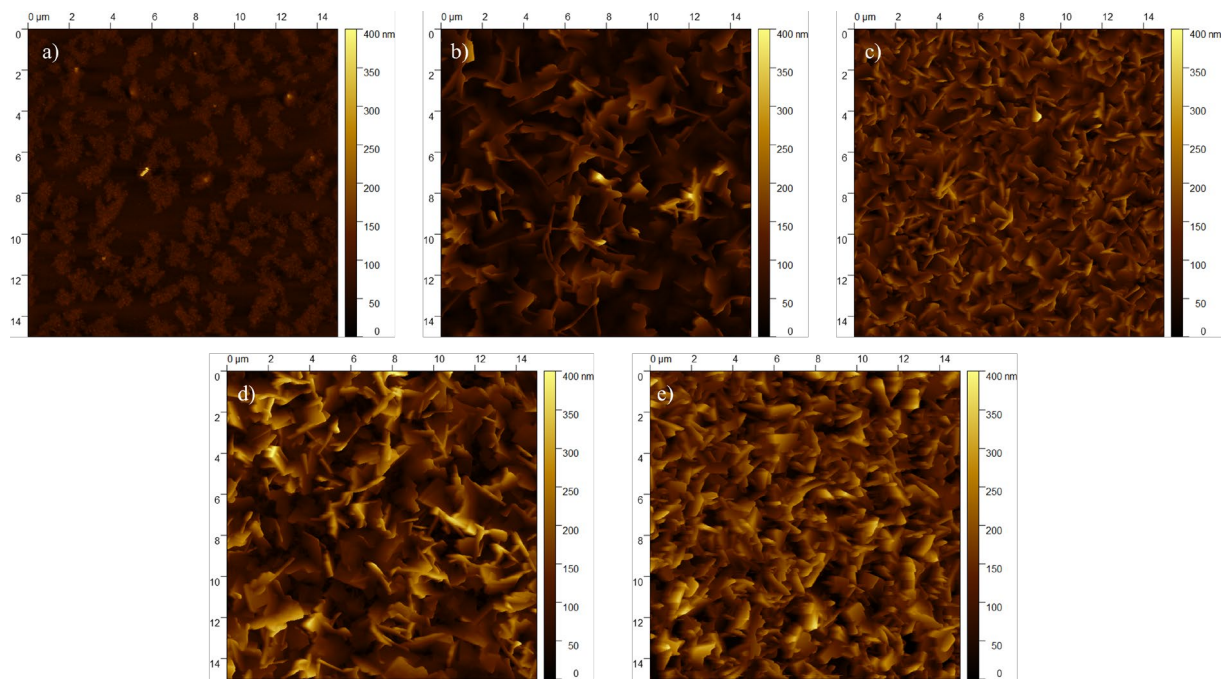

Figure S2. AFM images of O$_2$ plasma-treated samples: (a) 0 seconds, (b) 5 seconds, (c) 30 seconds, (d) 1 minute, (e) 5 minutes.

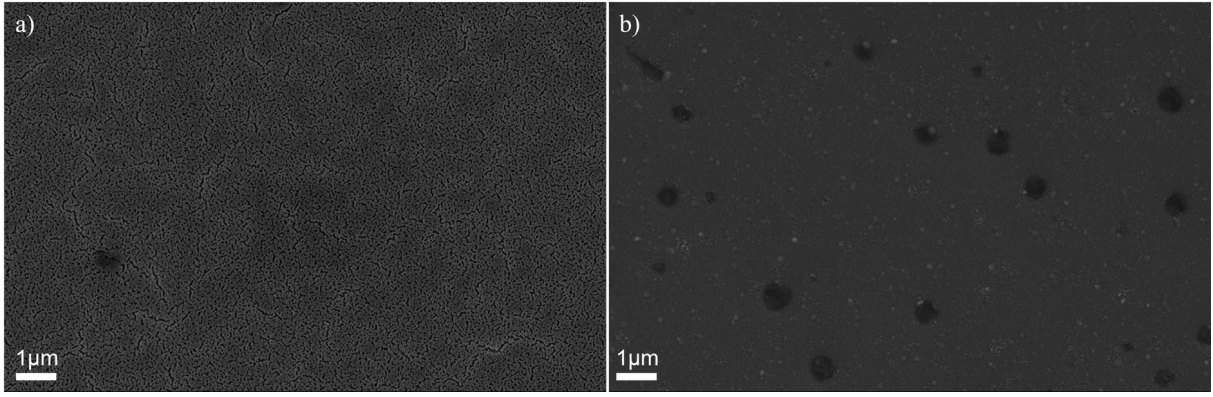

Figure S3. SEM images of (a) $O_2$ plasma-treated Ag and (b) $H_2O$ vapor pulse-treated Ag.

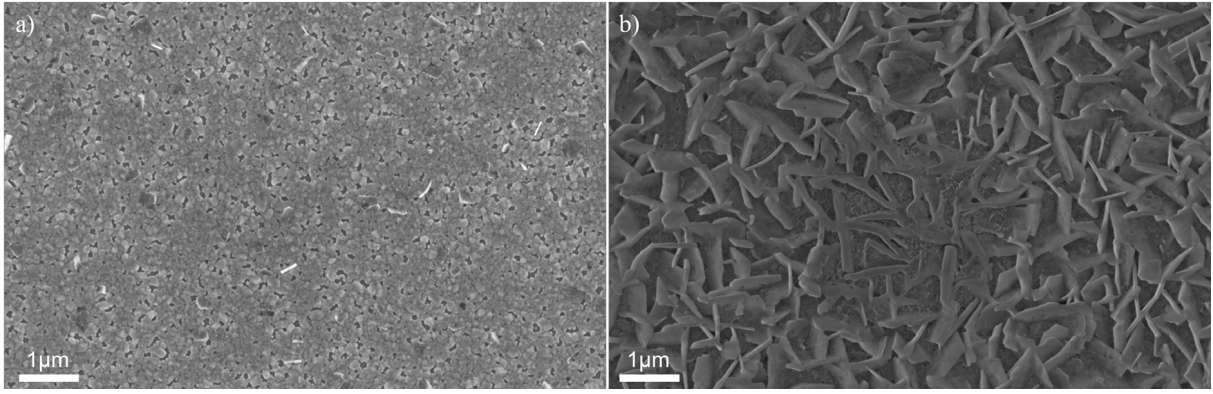

Figure S4. (a) SEM image of the H$_2$O vapor pulse-treated sample. (b) SEM image of the O$_2$ plasma-treated sample after one day of conversion.

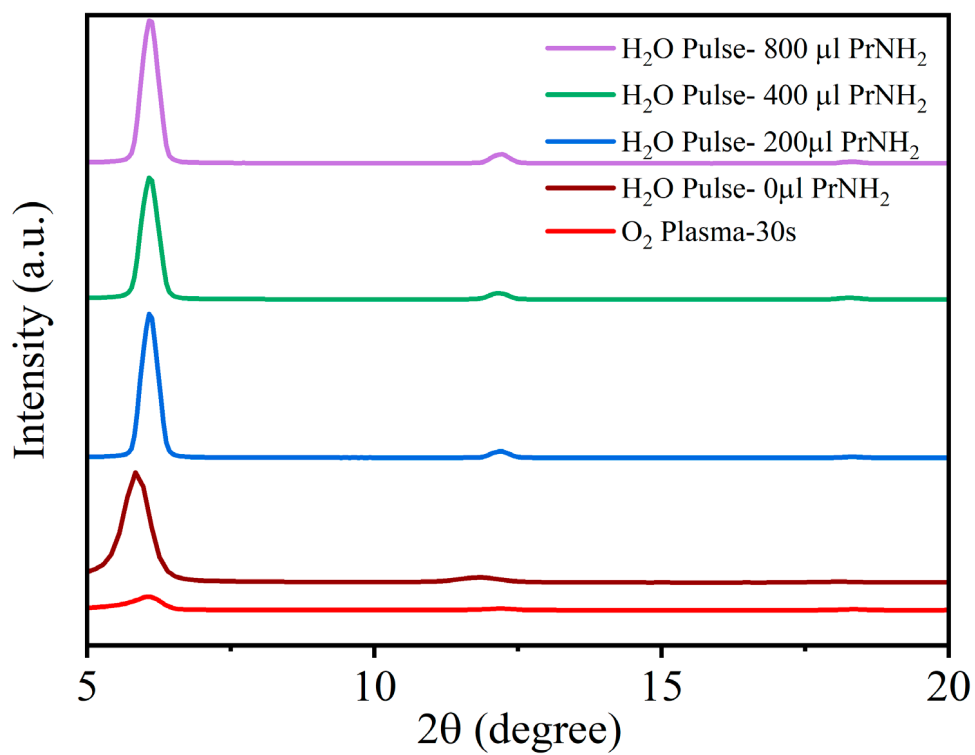

Figure S5. Comparison of XRD results for $O_2$ plasma-treated, $H_2O$ vapor pulse-treated, and $H_2O$ vapor pulse-treated samples modified with 200 μl, 400 μl, and 800 μl of the organic $PrNH_2$ ligand.

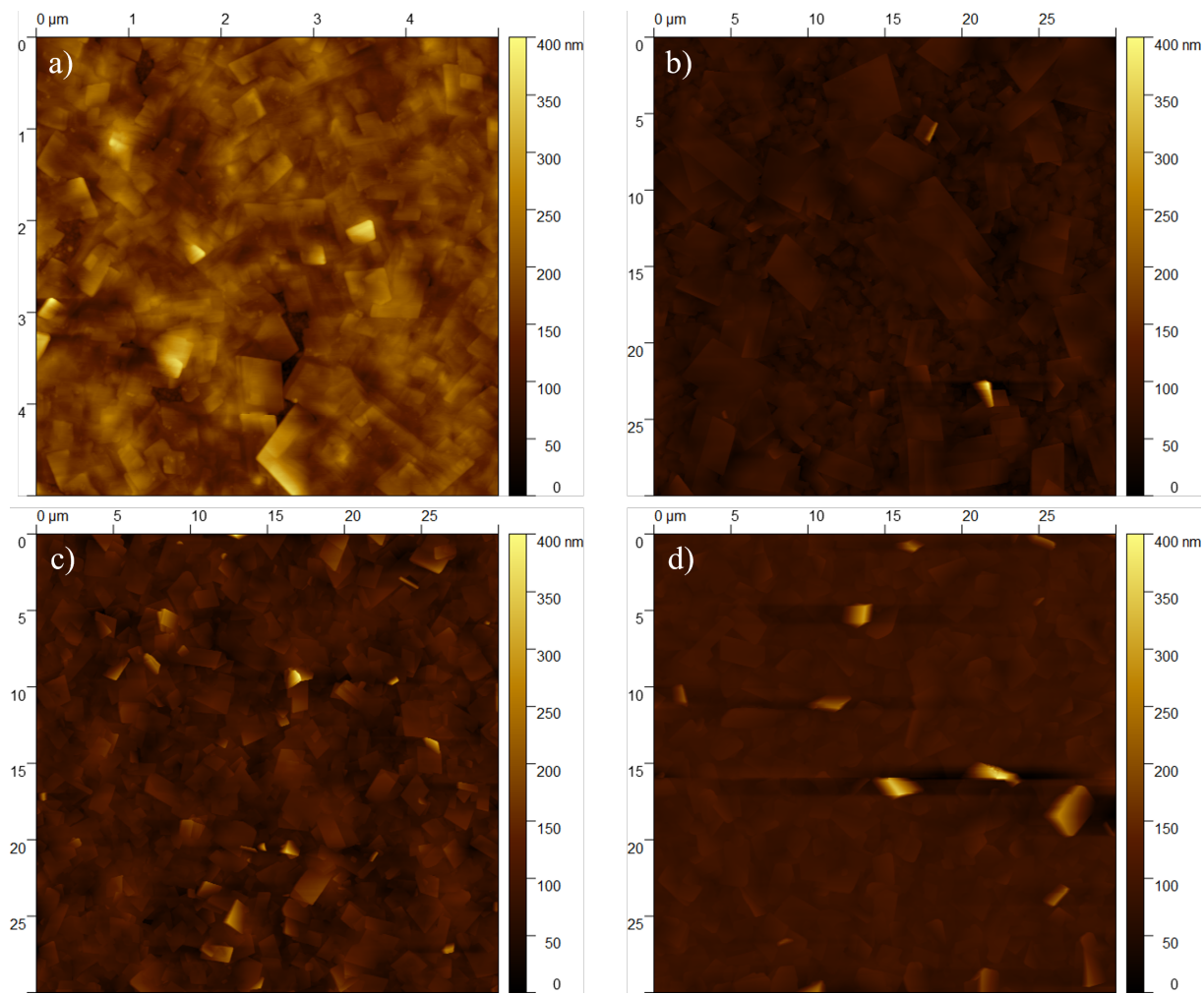

Figure S6. AFM images of H$_2$O vapor pulse-treated samples with (a) 0 µL PrNH$_2$, (b) 200 µL PrNH$_2$, (c) 400 µL PrNH$_2$, (d) 800 µL PrNH$_2$.

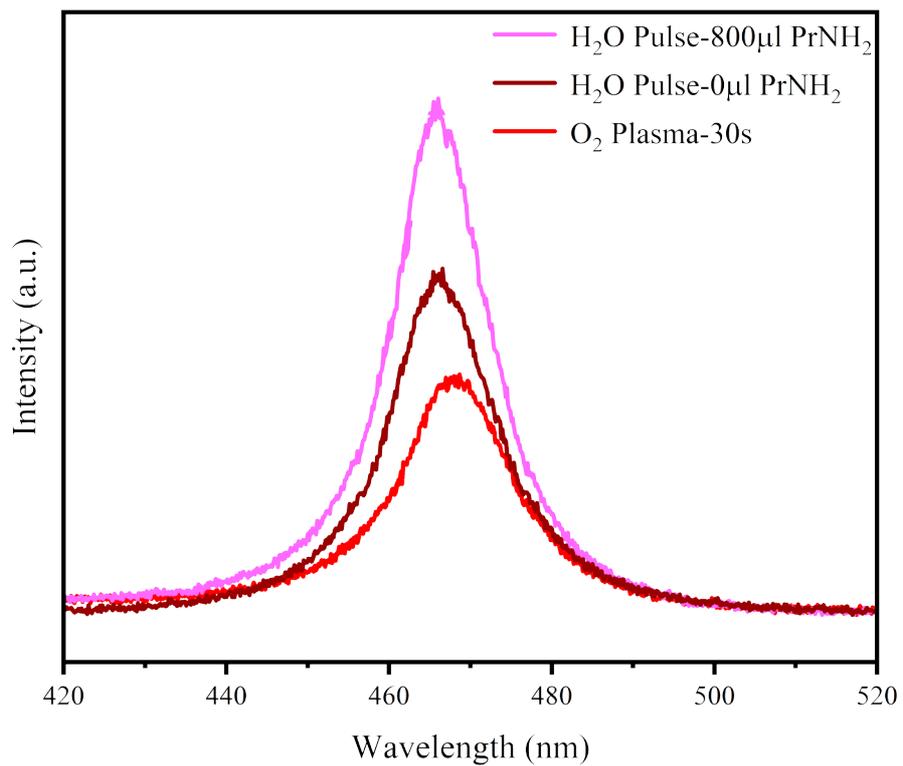

Figure S7. Comparison of photoluminescence for $O_2$ plasma-treated, $H_2O$ vapor pulse-treated, and $H_2O$ vapor pulse- treated samples modified with 800 µl of the organic $PrNH_2$ ligand.

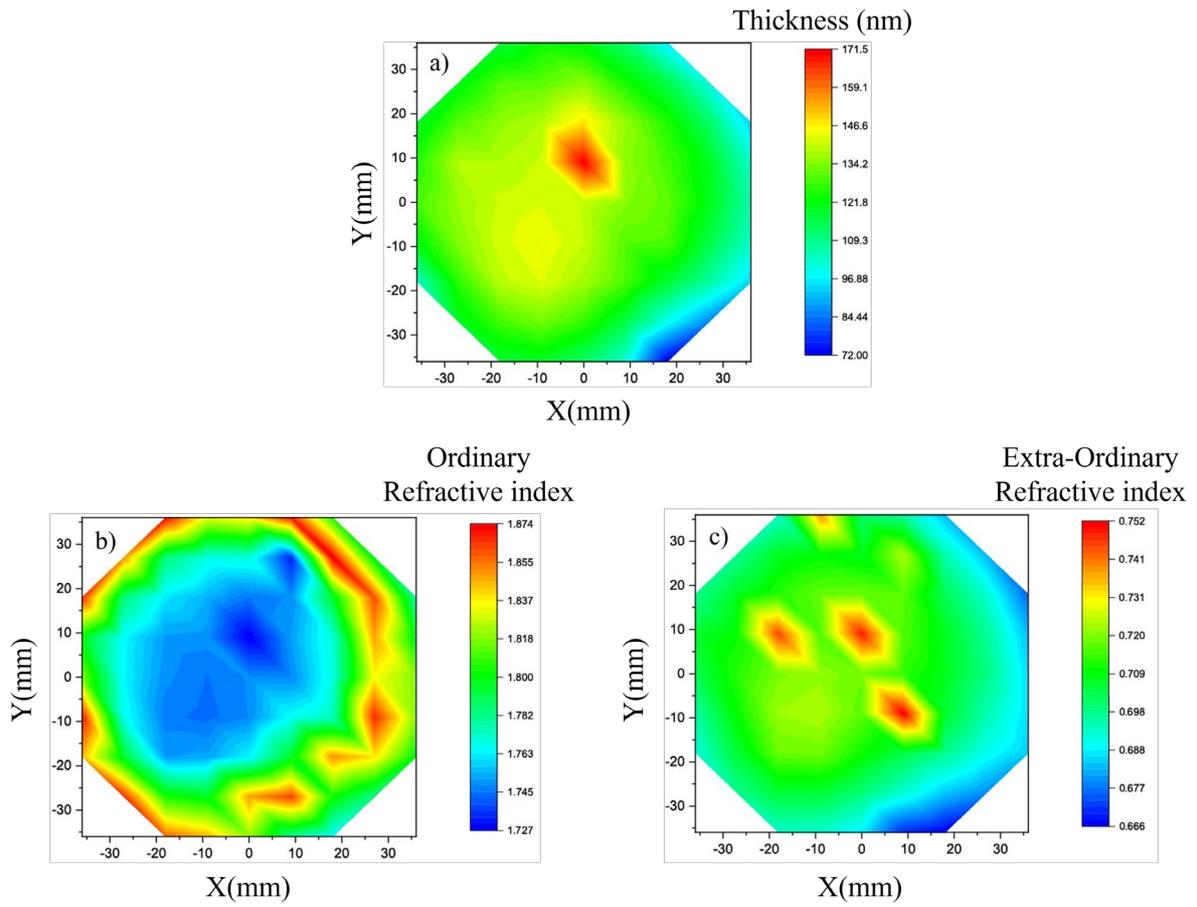

Figure S8. Spectroscopic ellipsometry mapping for the mithrene growth on the 100 mm wafer: (a) Thickness, (b) Ordinary refractive index, (c) Extraordinary refractive index.

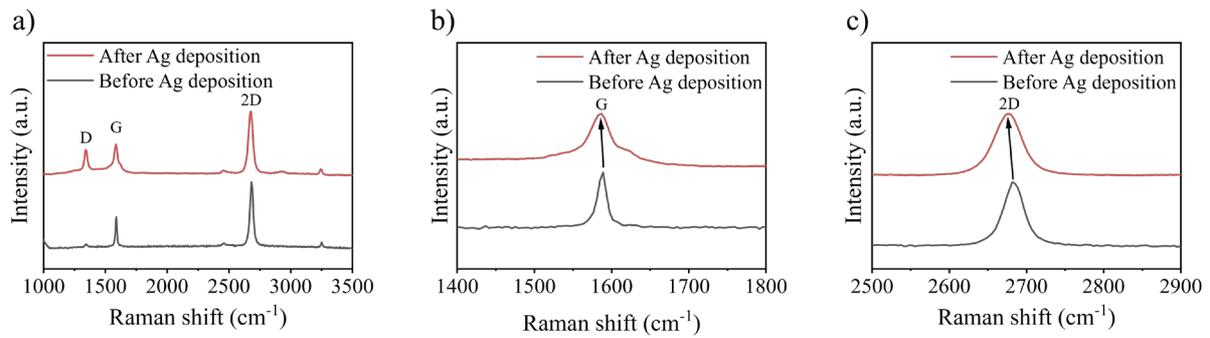

Figure S9. (a) Raman peaks of single-layer graphene before and after Ag deposition, obtained using an excitation wavelength of 532 nm, (b and c) Raman spectra showing the G peak and 2D peak, respectively. The black arrows serve as guidelines to indicate the downshift of the G and 2D peaks.

The Raman spectra in Figure S9a show the characteristic G and 2D bands at approximately 1587 cm$^{-1}$ and 2700 cm$^{-1}$, respectively, as well as the D band at approximately 1355 cm$^{-1}$. The data taken before Ag deposition (black) shows a higher intensity of the 2D peak than the G peak, and a very small D peak, indicating a good quality of the single-layer graphene. After silver deposition, a higher D peak appears, indicating more defects. The black arrows serve as guidelines to illustrate blueshifts of the G and 2D peaks due to Ag-induced doping (Figures S9b and c)[1].